\documentclass[%
 aip,
 amsmath,amssymb,
reprint,%
]{revtex4-1}

\usepackage{graphicx}
\usepackage{dcolumn}
\usepackage{bm}
\usepackage{xcolor}

\usepackage{multirow,units,array,notes2bib}
\newcommand{\lmat}{\left[\!\!\begin{array}{ccccccccc}}
\newcommand{\rmat}{\end{array}\!\!\right]}

\begin{document}

\preprint{AIP/123-QED}

\title[]{Toward Microphononic Circuits on Chip:  An Evaluation of Components based on High-Contrast Evanescent Confinement of Acoustic Waves}

\author{Yangyang Liu}
\email{yangyang.liu@colorado.edu}
\author{Nathan Dostart}
\affiliation{Department of Electrical, Computer, and Energy Engineering, University of Colorado, Boulder, CO 80309, USA}
\author{Milo\v{s} A. Popovi\'{c}}
\affiliation{Department of Electrical and Computer Engineering, Boston University, Boston, MA 02215, USA}

\date{\today}%

\begin{abstract}
We investigate the prospects for micron-scale acoustic wave components and circuits on chip in solid planar structures that do not require suspension.  We leverage evanescent guiding of acoustic waves by high slowness contrast materials readily available in silicon complementary metal-oxide semiconductor (CMOS) processes.  High slowness contrast provides strong confinement of GHz frequency acoustic fields in micron-scale structures.  We address the fundamental implications of intrinsic material and radiation losses on operating frequency, bandwidth, device size and as a result practicality of multi-element microphononic circuits based on solid embedded waveguides.  We show that a family of acoustic components based on evanescently guided acoustic waves, including waveguide bends, evanescent couplers, Y-splitters, and acoustic-wave microring resonators, can be realized in compact, micron-scale structures, and provide basic scaling and performance arguments for these components based on material properties and simulations.  We further find that wave propagation losses are expected to permit high quality factor (Q), narrowband resonators and propagation lengths allowing delay lines and the coupling or cascading of multiple components to form functional circuits, of potential utility in guided acoustic signal processing on chip.  We also address and simulate bends and radiation loss, 
providing insight into routing and resonators.  Such circuits could be monolithically integrated with electronic and photonic circuits on a single chip with expanded capabilities.\end{abstract}

\maketitle

\section{Introduction}
With the tremendous advances in modern lithography and high-resolution nanofabrication that were driven by the electronic integrated circuit (IC) industry, micron-scale photonic circuits, including silicon photonic circuits, have emerged over the past five decades, as a new chip technology showing substantial promise to enable many applications and provide performance superior to electronics (particularly for communication), as well as to enhance CMOS technology itself \cite{kaminow2008optical,sun2015single}.  High optical refractive index contrast between core and cladding materials in planar photonic waveguide geometries has allowed small, wavelength-scale microphotonic and optoelectronic components to be integrated in complex circuits and systems on a chip, providing capabilities for a wide range of compact high performance applications in communication, sensing and information processing \cite{popovic2008theory}.  Since acoustic and optical (electromagnetic) waves share many mathematical similarities \cite{auld1990acoustic}, an analogous high slowness contrast acoustic wave confinement scheme could bring the same possibilities to acoustics to enable microphononic circuits, where the manipulation of signals in the acoustic domain can maximally benefit from the advancing integration technology and wavelength scale confinement.  A key benefit is that evanescent confinement provides for efficient coupling between circuit elements such as waveguides and resonators.  The viability of such a scheme will be determined by the achievable device size scale and propagation losses -- which will determine the complexity (how far waves can propagate, i.e. through how many components), and the bandwidth (i.e. how long acoustic waves can spend in the circuit without dissipating).  

Coherent phonons travel at much slower speeds than photons and can directly interact with radiofrequency or optical signals\cite{fang2016optical}, making chip-scale acoustics or microphononics an interesting domain for researchers in phononics, optomechanics and micro-/nano-electro-mechanical systems (MEMS/NEMS). A wide range of systems have recently been explored, where acoustic waves interact with electronics and photonics in chip-scale integrated platforms to enable optical delay lines \cite{safavi2011electromagnetically,fang2016optical}, narrow-linewidth RF photonic filters \cite{shin2015control}, photonic-phononic memory \cite{merklein2016chip} , frequency locking of micromechanical oscillators \cite{shah2015master}, systems in the quantum mechanical ground state \cite{chan2011laser}, and  on-chip optomechanical signal detection \cite{sun2014monolithically}. The ability to confine and guide acoustic waves not only enables the construction of individual devices, but may also provide a means to connect them to form a complete on-chip acoustic circuitry.

Early investigation into acoustic waveguides proposed various geometries including flat overlay waveguides with thin metal strips deposited on a substrate, topological waveguides that confine acoustic fields to a ridge or wedge in a homogeneous material, and in-diffused waveguides that are locally implanted with metals to create a weak impedance contrast \cite{oliner1976waveguides}. 
Later developments in surface acoustic wave (SAW) and bulk acoustic wave (BAW) technologies make use of acoustic fields vertically bound to a free material surface or between air and an acoustic reflector, and have enabled commercial applications in radar and communication systems, radio/intermediate frequency signal processing, and chemical/biological sensing \cite{ruppel2000advances,campbell1989applications}, despite lack of lateral confinement that allows sharp bends. Acoustic confinement in off-chip cylindrical waveguides has also been investigated, e.g. in the context of Brillouin scattering in low-contrast silica optical fibers \cite{shelby1985guided} and analytically in GaAs-AlAs quantum wires  \cite{nishiguchi1994guided}.
Modern fabrication technologies have enabled the construction of planar structures with an inhomogeneous distribution of materials in the cross-section.  
Chip-scale acoustic devices so far still heavily rely upon 1) air suspension, where discontinuation of solid material reflects acoustic waves at free boundaries \cite{shin2013tailorable},  
and 2) phononic crystal structures, where engineered slowness mismatch in a periodic composite material provides wave confinement \cite{marathe2014resonant}, or a combination of these two mechanisms \cite{hatanaka2014phonon}.
Recently, resurgent interest in optomechanics and stimulated brillouin scattering in on-chip devices has led to investigations of novel acoustic waveguiding geometries for strong light-sound coupling, including geometrical \cite{sarabalis2016guided} and material slowness contrast based (evanescent) \cite{poulton2012design,poulton2013acoustic} confinement.

In this paper, we investigate the prospects of evanescent guiding for ultracompact acoustic components on chip in silicon-based CMOS materials.  We look at waveguides and resonators, and investigate the limitations expected to be imposed by intrinsic material loss on device lengths and resonator Q's, and show in simulation designs of several evanescent-confinement based components, including a directional coupler, Y-splitter and acoustic microring resonator.
Evanescent confinement means that the structures need not be air-suspended, making them suitable beyond MEMS-like processes and potentially integrable with electronics and photonics in mainstream advanced CMOS \cite{sun2015single}.
These geometries could also be readily employed to add lateral confinement to conventional SAW devices, in MEMS circuits to provide evanescent coupling as an alternative to coupling springs, and in optomechanics to permit simultaneous and controlled guiding of both acoustic and optical waves. 

In the remainder of the paper, we first study the guidance properties and confinement strength of evanscently-confined waveguides in a high slowness contrast scenario.  Second, we evaluate the effect of material losses in typical silicon CMOS materials, as well as the effect of radiation loss in curved waveguides where the phase velocity exceeds the local speed of sound.  We provide some scaled plots that provide an idea of the parameter space where useful waveguides and resonators can be designed.  Last, we simulate a few components including an evanescent directional power coupler, an acoustic Y-splitter, and a microring resonator, showing 10$\mu$m scale components in all cases.  Our study suggests that evanescent confinement can support complex multi-element phononic guided-wave and resonator circuits on chip, at frequencies from below 1\,GHz to 10's of GHz where they may have the potential to offer better performance than photonic circuits.

\section{Acoustic Wave Equation and Slowness Curves}
Guiding of acoustic waves has marked similarities to that of optical waves.  
At a key time in the development for acoustic devices in the 1970's, Auld reformulated and mapped the acoustic equations and conservation laws by following the example of Maxwell's equations \cite{auld1990acoustic}.  
In refraction of both electromagnetic and acoustic waves at a material interface, the tangential component of the wave vector is conserved, as a result of continuity of tangential electric field and continuity of particle velocity, respectively. Evanescent guiding between two materials requires existence of a critical angle, beyond which propagating waves are confined in the material with slower wave speed. In optics, refractive index is used to describe the ratio between speed of light in vacuum and in the material. For acoustic waves, since there is not a common reference, slowness, defined to be the inverse of wave speed in the material, is used to the same effect as optical index of refraction.

\begin{figure}[!htbp]
\includegraphics[width=.5\textwidth]{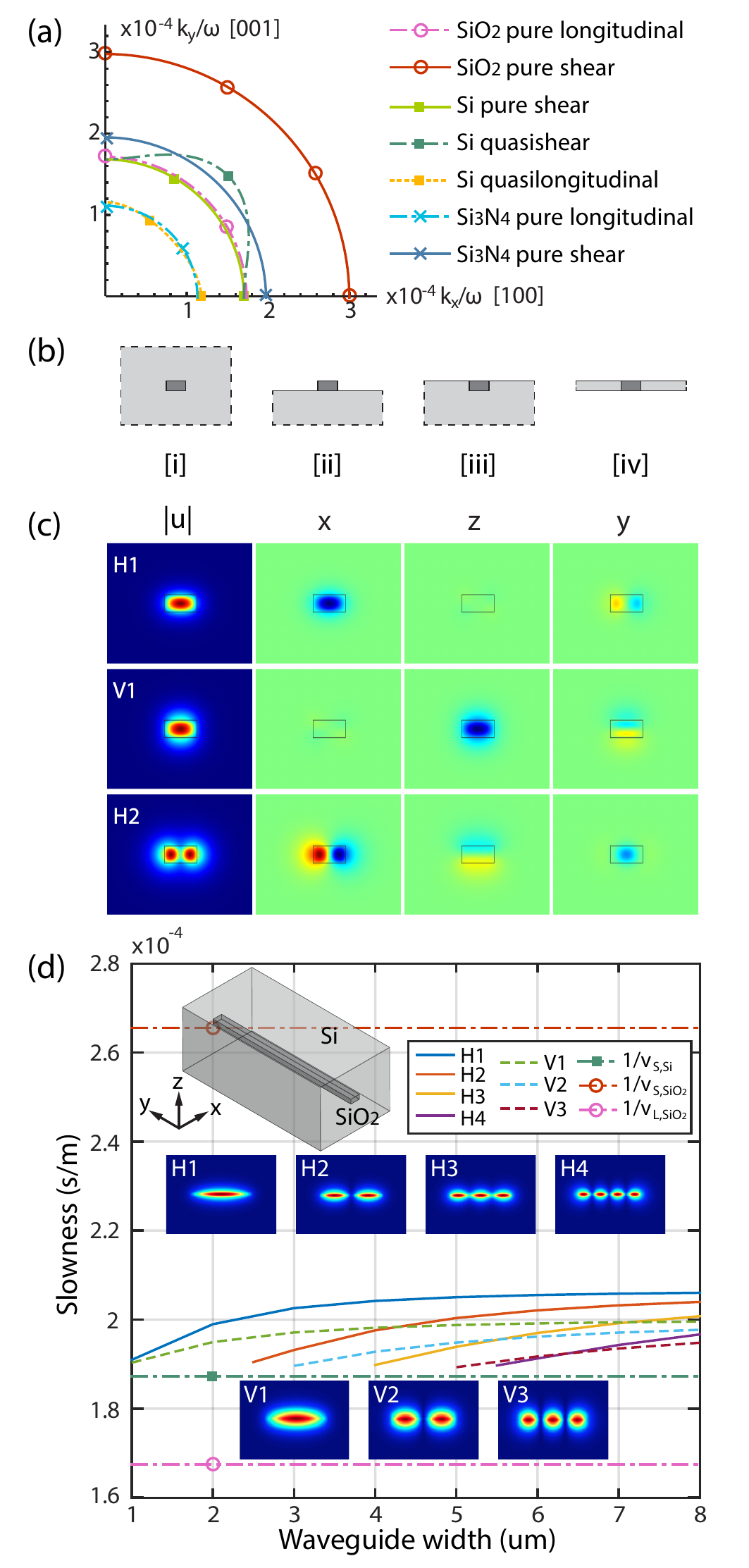}
\caption{(a) Slowness curves of acoustic plane waves propagation along crystal faces in bulk Si, SiO$_2$ and SiN. (b) Possible geometries for high-contrast dielectric acoustic waveguides: [i] fully-cladded [ii] half-cladded [iii] half-cladded, embedded [iv] side-cladded, membrane. (c) Total displacement $|u|$ (blue/red represents min/max displacement), and horizontal $u_x$, vertical $u_z$ and longitudinal $u_y$ components (blue/red represents negative/positive displacement) of the first three horizontally (H) and vertically (V) polarized eigenmodes in a fully cladded rectangular waveguide with a $1\mu$m$\times 2\mu$m SiO$_2$ core surrounded by Si cladding.  (d) Slowness ($1/v_p$) v.s. waveguide core width for fixed waveguide height of $0.5\mu$m and wavelength $\lambda = 1.5\mu$m.}
\label{fig:modes}
\end{figure}

In this section, we derive the slowness curves for the propagation of plane waves in relevant core and cladding materials and find the conditions for evanescently-confined guided waves.  
At the interface between two materials with different stiffnesses and densities, acoustic waves propagating in one material can, under certain conditions, refract into only evanescent waves in the other material. This acoustic version of total internal reflection can confine acoustic fields in a slow (analogous to a high refractive index for light) core material surrounded by a fast cladding. In addition to waveguides, traveling and standing wave resonators and other functional components can be designed using the same configuration. Propagation of acoustic waves in solids is governed by the acoustic wave equation
\begin{equation}
\nabla \cdot \mathbf c:\nabla_s \mathbf v=\rho\,\partial_t^2\mathbf v-\partial_t\mathbf F
\label{eqn:wave}
\end{equation}
where $\mathbf v$ is the velocity field, $\nabla_s$ is the symmetric gradient operator, $\mathbf c$ and $\rho$ are the material stiffness tensor and mass density, and $\mathbf F$ is external driving force density. 
In a source free case ($\mathbf F=0$), eigenstate waves are supported with displacement field proportional to $\mathbf u \sim\exp[i(\omega t-k\mathbf{\hat l}\cdot\mathbf r)]$ and propagate along
$\mathbf{\hat l} = \mathbf{\hat x}l_x+\mathbf{\hat y}l_y+ \mathbf{\hat z}l_z $. 
Equation~\ref{eqn:wave} can be rearranged based on an assumed plane-wave solution to derive the Christoffel equation \cite{auld1990acoustic}
\begin{equation}
\left[k^2\Gamma_{ij}-\rho\omega^2\delta_{ij}\right][v_j]=0
\label{eqn:christoffel}
\end{equation}
where 
$\Gamma_{ij}=l_{iK}c_{KL}l_{Lj}$
is the Christoffel matrix, containing elements of the $\nabla \cdot \mathbf c:\nabla_s$ operator applied to the assumed plane wave solution.  Setting the determinant of Eq.~\ref{eqn:christoffel} to zero,
\begin{equation}
\Omega(\omega,k_x,k_y,k_z)=\left|k^2\Gamma_{ij}(l_x,l_y,l_z)-\rho\omega^2\delta_{ij}\right|=0
\label{eqn:dispersion}
\end{equation}
a dispersion relation in the form of slowness surfaces $1/v_p\equiv k/\omega$ can be fully solved for plane wave propagation in a homogeneous material.
Figure~\ref{fig:modes}(a) shows the slowness curves for plane waves propagating along crystal faces in three materials commonly used in a silicon platform, Si, SiO$_2$ and SiN. Silicon, like other crystalline materials in general, is acoustically anisotropic and tri-refringent,  yielding three distinct slowness curves as solutions to Eq.~\ref{eqn:dispersion}. SiO$_2$ and SiN are amorphous and isotropic.  In this case, the two shear polarizations become degenerate.

\section{Guiding Acoustic Waves in Solids}

In an inhomogenous material geometry, scattering at a material boundary requires that the particle displacement and normal stress fields be continuous.  For example, at a planar interface at $z = 0$ (normal position $\mathbf r_\perp \equiv \hat{z} z = 0$), the tangential wave vector $\mathbf k_\parallel \equiv \hat{x}k_x + \hat{y}k_y$ must be equal in both materials to give the same spatial field dependence $\exp(i\mathbf{k}_\parallel\cdot\mathbf r_\parallel)$ that avoids discontinuities.  In the slowness picture, when a wave is incident on the interface from one side, this means that the tangential component of $1/\mathbf{v}_p = \mathbf k/\omega$ must be conserved across the interface as a corresponding wave is excited in transmission on the other side of the interface.  If the interface undulation is slower than the wave in the second, cladding material, then the ``transmitted'' wave is evanescent with an imaginary $\mathbf{k}_\perp$, as in the familiar case of optical waveguides.

However, since materials are in general trirefringent for acoustic propagation, with three slowness curves,
for one longitudinal and two shear/transverse polarizations (amorphous materials are isotropic with degenerate shear modes), acoustic confinement is more complex to achieve than optical confinement. Depending on the relation between wave speeds in materials, up to five critical angles can exist, $\theta_\text{cr} = \sin^{-1}(\frac{v_i}{v_r})$,  where $v_i$ and $v_r$ are the velocities of the incident and refracted waves and $v_i<v_r$. These critical angles are associated with reflected waves of other polarizations in the incident material, and refracted waves in the other material. Beyond all possible critical angles the incident wave undergoes total internal reflection. For a cladded structure to be evanescently confining, the slowness of its waveguide modes must fully enclose the slowness curves of all polarizations in the cladding, to prevent scattering into propagating bulk modes.
%
%
In a rectangular waveguide with Si cladding and SiO$_2$ core, 
sidewalls couple shear and longitudinal plane waves into hybrid modes, among which the ones with primarily shear components have slower speeds than the fastest polarization in the cladding and can thus be fully confined in the SiO$_2$ core.  
Figure~\ref{fig:modes}(b) illustrates four possible configurations for high-contrast dielectric acoustic waveguides utilizing evanescent confinement between two materials and free air interfaces that can potentially suit different applications and fabrication platforms. The fully-cladded waveguide (Fig.~\ref{fig:modes}(b)[i]) consists of a slow core surrounded by a fast cladding, drawing a direct analogy to cladded rectangular optical waveguides.  Figure~\ref{fig:modes}(b)[ii] and [iii] are half-cladded waveguides with the top surface of the core exposed to air. The embedded version (Fig.~\ref{fig:modes}(b)[iii]) has a planar top surface that could further interface to transducers, such as patterned metal electrodes.  Side-cladded waveguides on an air-suspended membrane (Fig.~\ref{fig:modes}(b)[iv]) provide an alternative to phononic crystal waveguides for systems without a compatible layer stack to evanescently confine acoustic waves in the vertical direction.

As a simple example to illustrate the behavior of evanescent mode acoustic waveguides, Fig.~\ref{fig:modes}(c) shows the mode field distributions of a fully-cladded rectangular waveguide with SiO$_2$ core and Si cladding simulated using COMSOL Multiphysics \cite{multiphysics2012comsol}.
A 3D section of the waveguide was simulated using the Solid Mechanics module, with Perfectly Matched Layer (PML) boundary conditions applied on the lateral faces and Floquet boundary condition on the longitudinal faces for a given propagation constant.
Guided modes split into two categories, horizontal (H) modes with a strong field component in the transverse plane along the horizontal edges of the core cross-section and vertical (V) modes along the vertical edges. 
Figure~\ref{fig:modes}(d) shows the slowness curves of the first few horizontal and vertical modes versus waveguide width for a fixed height of $0.5\mu$m. All guided modes are below the shear wave slowness in the SiO$_2$ core, but above that in the cladding, at which point waves in the core start refracting into the cladding, losing evanescent confinement at the interface. The width of an acoustic waveguide can be designed through numerical simulation to support only the lowest order horizontal and vertical modes, which approach degeneracy at small waveguide widths, and thus there is no apparent cutoff for the fundamental vertically polarized (V1) mode.

\begin{figure}[t!]
\includegraphics[width=.5\textwidth,trim=0 0 0 0,clip]{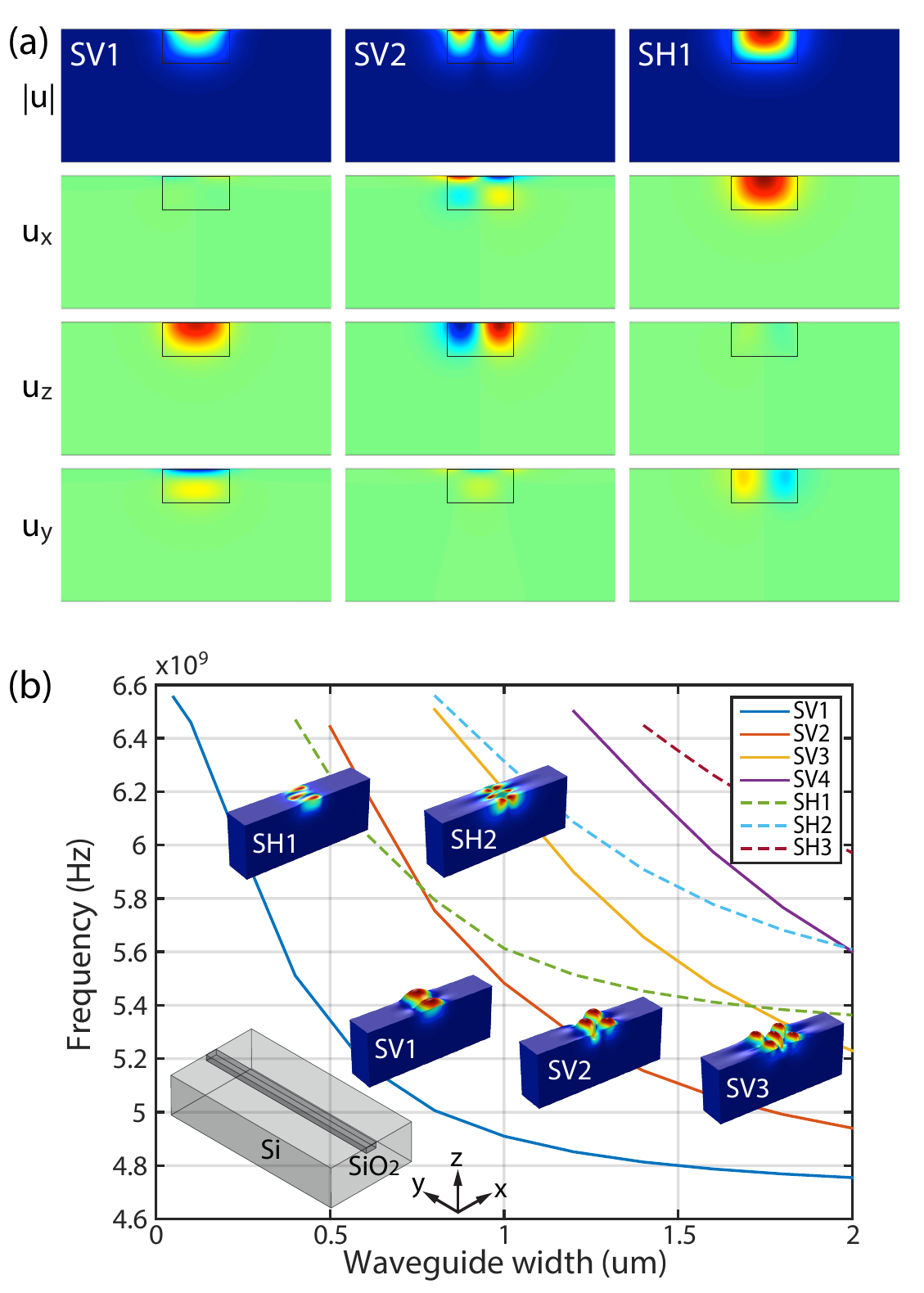}
\caption{(a): Total displacement $|\mathbf u|$ (blue/red represents min/max displacement), and horizontal $u_x$, vertical $u_z$ and longitudinal $u_y$ components of the first three vertically (SV) and horizontally (SH) polarized surface modes (blue/red represents negative/positive displacement) of a half-cladded waveguide with a $0.5\mu$m$\times1\mu$m Si core embedded in SiO$_2$ cladding. (2) Slowness (1/$v_p$) v.s. waveguide core width for fixed waveguide height of 0.5 $\mu$m and wavelength $\lambda$= 0.75 $\mu$m.}
\label{fig:halfclad}
\end{figure}

Combining evanescent confinement and free air interfaces, the embedded half-cladded geometry, Fig.~\ref{fig:modes}(b)[iii], can be a more practical configuration in terms of fabrication, and it provides better access to the mode fields that are tightly concentrated on the exposed top surface of the core and leverage surface wave components.  Figure~\ref{fig:halfclad}(a) shows the field components of the first three vertically (SV) and horizontally polarized (SH) surface modes. The SV modes have a dominant vertical shear component ($u_z$), which spreads across the core with field maxima on the top surface; the shear horizontal ($u_x$) and longitudinal ($u_y$) components are evanescently confined to the top surface, resembling Love waves in a half-cladded slab \cite{auld1990acoustic}. The SH waves have a dominant shear horizontal ($u_x$) component. The half-cladded acoustic waveguides could be fabricated directly into a silicon substrate using localized thermal oxidation defined by a hardmask.  Alternative fabrication methods include chemical vapor deposition of SiO$_2$ on a pre-patterned silicon trench followed by chemical mechanical planarization (CMP), or a different fast material such as SiN can be used as a substrate underneath the Si/SiO$_2$ layer to provide a bottom cladding, and the oxide waveguide can be patterned from a device layer using regular scanning electron beam or optical lithography.  This waveguide geometry has tight lateral confinement that allows for sharp bends with low radiation loss, enabling ultra-compact components, and high concentration of fields on the top surface, making it a versatile candidate to allow interfacing acoustic waves to other systems with different signal carrying physics, such as electronic circuits or guided light waves.

Our ultimate goal in considering high slowness contrast evanescent confinement is to construct chip-scale phononic components and coupled-element circuits that can enable a richer signal processing capability.  
High slowness contrast enables compact components on the order of 10\,$\mu$m, which will be discussed in more detail in Sec.~\ref{sec:components}.  However, the practicability of a phononic circuitry such as that described here requires that signal propagation lengths in waveguides are large enough to traverse a few components, and that excitation lifetimes in resonators are long enough to process (e.g. filter or delay) relevant bandwidth signals and couple the energy faster to the next circuit element than to the radiation loss mechanisms.  Therefore, in the following, we first consider acoustic loss mechanisms and the bounds they place on performance of waveguides and resonators in order to evaluate the viability of high-contrast microphononic circuits.  We consider intrinsic material losses in the next section, and then radiation loss -- a generalized anchor loss mechanism -- in the following section on device design.

\section{Impact of Material Intrinsic Losses}

A key consideration in understanding the utility, and range of applicability, of these acoustic waveguides is loss.  In this paper we consider two mechanisms: material intrinsic loss, dealt with in this section, and acoustic radiation loss, a fundamental loss mechanism which occurs due to bending of otherwise lossless waveguides, discussed in the next section because it is associated with device design.  For conventional on-chip acoustic devices, energy dissipation is mainly caused by air damping, anchor loss and intrinsic material losses  \cite{lobontiu2014dynamics,wu2013effect}.  Air damping includes a few different mechanisms and is often dominated by squeeze-film damping for suspended structures with small air gaps between a vibrating film and a stationary substrate \cite{brotz2004damping,wu2013effect}. Anchor loss is caused by acoustic radiation into the substrate through attachments such as pedestals and spokes \cite{liu2014high,liu2014ultra} that provide mechanical support for suspension. Having acoustic waves fully confined in solids evanescently (without air suspension) exempts this type of devices from squeeze-film damping and from conventional anchor loss (radiation into the cladding, in straight sections), but intrinsic material losses still impose a limit on the frequency range where low loss waveguides and resonant cavities with high quality factors can be achieved. Hence they are addressed first in the context of their impact on wavelength scale devices and circuits.

The two main intrinsic loss mechanisms for acoustic devices are thermoelastic dissipation (TED) and phonon-phonon interaction associated dissipation (PPD), which both result from coupling between the acoustic field and thermal phonons in solids, at different time and length scales \cite{braginsky1985systems,duwel2011thermal,maris2012interaction,chandorkar2008limits,ayazi2011energy}.  The total material intrinsic loss limited quality factor $Q_\text{intrinsic}$ is given by $Q_\text{intrinsic} ^{-1}=Q_\text{TED}^{-1} +Q_\text{PPD}^{-1} $, 
where ${Q_\text{TED}}$ and ${Q_\text{PPD}}$ are Q limits due to TED and PPD respectively.    

TED describes the energy dissipation associated with coupling between a strain field and a temperature gradient through irreversible heat flow (absent in volume preserving pure shear waves). TED is dependent on the specific geometry of an acoustic device, and can be minimized in design by reducing the overlap of the strain field induced by the acoustic wave and the heat diffusion eigenmodes of the system \cite{chandorkar2008limits}, but the bulk limit still provides a general measure of feasibility of materials for acoustic wave guiding. Landau and Lifshitz calculated the attenuation coefficient for longitudinal waves in amorphous (isotropic) solids, which also gives an order of magnitude estimate for anisotropic crystals \cite{landau1959course}. An equivalent expression for $Q_\text{TED}$ at angular frequency $\omega$ and temperature $T$ is \cite{braginsky1985systems,chandorkar2008limits,ayazi2011energy} 
\begin{equation}
Q_\text{TED} = \dfrac{9C_v^2}{\omega \kappa T\beta^2\rho}\label{eqn:TED}
\end{equation}
where  $C_v$, $\kappa$, $\beta$ and $\rho$ are the volumetric heat capacity, thermal conductivity, thermal expansion coefficient and mass density.  

PPD describes interactions between incident acoustic waves and thermally excited phonons that happen at shorter time and length scales than TED.  Akhieser considered this loss mechanism in the low frequency regime $\omega \tau\ll1$ (Akhieser regime), where $\tau$ is the lifetime of the thermal phonons, and used the Boltzmann equation to derive the attenuation by calculating the increase in entropy due to collisions between thermal phonons \cite{akhiezer1939sound}.  Woodruff and Ehrenreich and others further developed this theory for $\omega\tau>1$ in the regime $\omega\ll k_BT/\hbar$ \cite{woodruff1961absorption}. A simplified expression for $Q_\text{PPD}$ in the extended frequency range $\omega\ll k_BT/\hbar$ is \cite{maris2012interaction,li2016attenuation}
\begin{equation}
Q_\text{PPD} = \dfrac{1+\omega^2\tau^2}{\omega\tau}\dfrac{\rho v_g^2}{C_vT}\left(\langle\gamma^2\rangle-\langle\gamma\rangle^2\right)^{-1}
\label{eqn:PPD}
\end{equation}
where $v_g$ is the group velocity of the acoustic wave, related to the mode (or phase) velocity $v_p$ by $1/v_g \equiv \frac\partial{\partial \omega} (\omega/v_p(\omega))$, and $\gamma$ is the phonon mode Gr{\"u}neisen parameter.  Here, neglecting dispersion, $v_g$ is taken to be approximately equal to the phase velocity of the acoustic wave $v_p=\omega/k$, and $\tau$ is estimated using the kinetic relation
$\kappa=\frac13\tau C_vV_D^2$ and $3V_D^{-3} = V_l^{-3}+2V_t^{-3}$, where
$V_D$, $V_l$ and $V_t$ are the Debye, longitudinal and transverse wave velocities \cite{braginsky1985systems,maris2012interaction}. $\langle\rangle$ indicates averaging over interacting thermal phonon modes \cite{anderson2000gruneisen,li2016attenuation,nava1976akhiezer}.

\begin{figure}[!hbtp]
\includegraphics[width=.5\textwidth,trim=0 0 0 0,clip]{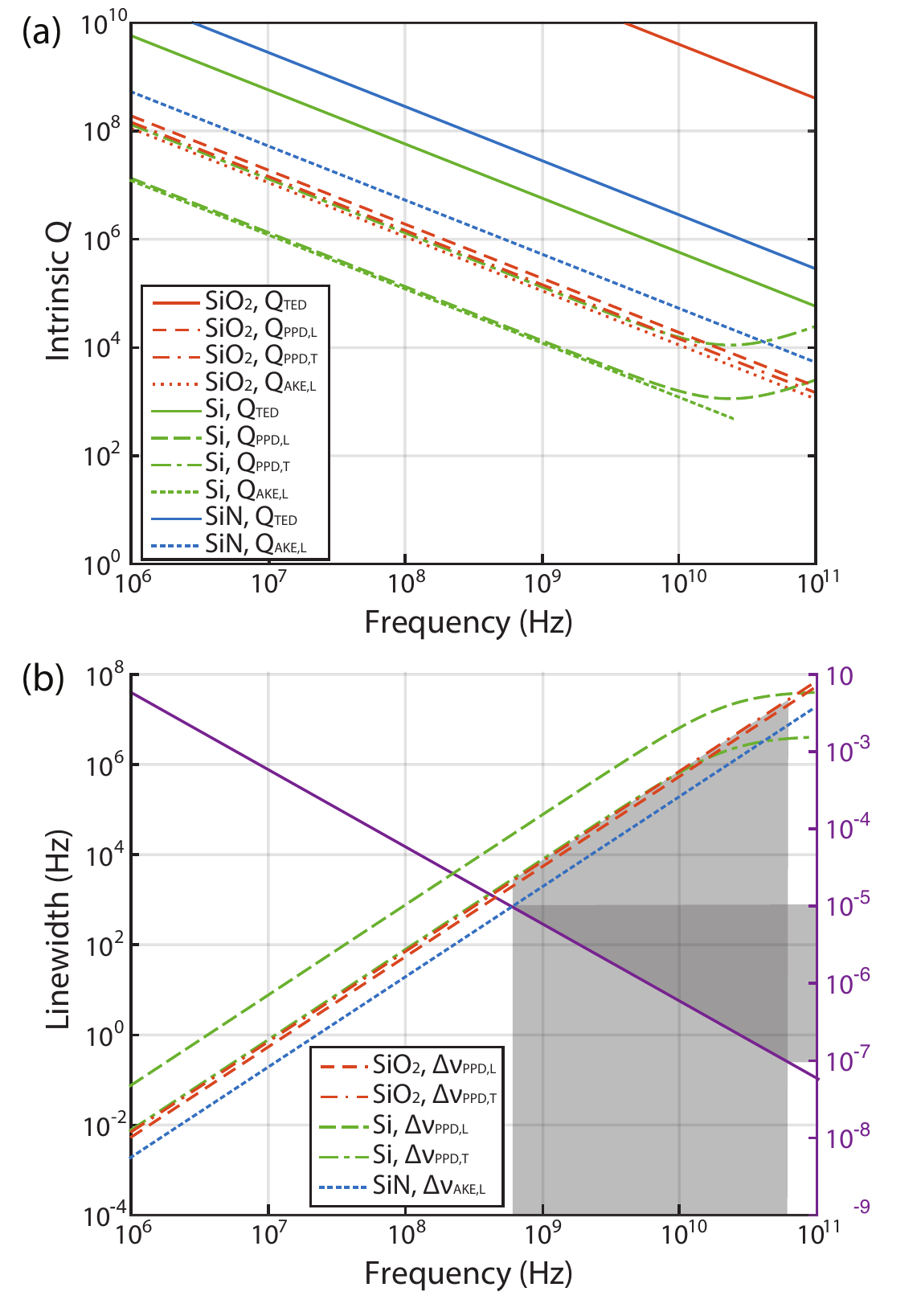}
\caption{(a) Q limits due to thermoelastic (solid lines) and phonon-phonon dissipations (dashed lines) for SiO$_2$, Si and SiN. (b) Intrinsic linewidths corresponding to the Q limits (dashed/dotted lines) and characteristic wavelength (pure shear wave in bulk Si, solid line) scaling versus frequency.}
\label{fig:Qlimit}
\end{figure}

\onecolumngrid

\begin{table}[h]
\smallskip\noindent
\resizebox{\linewidth}{!}{%
\begin{tabular}{|c|cccccccccccc|}
\hline
\multirow{3}{*}{Parameter}&\multirow{2}{*}{density}&specific&thermal&thermal&\multicolumn{2}{c}{wave speed}&Debye&\multicolumn{4}{c}{Gr\"{u}neisen parameters}&phonon\\
&&heat capacity&conductivity&expansion&long.&trans.& temperature&long.&trans.&long.&trans.&lifetime\\
&(kg/m$^3$)&(J/kg$\cdot$K)&(W/m$\cdot$K)&($\times 10^{-6}$K$^{-1}$)&(m/s)&(m/s)&(K)&&&&&($\times10^{-12}$s)\\
\hline
Symbol&$\rho$&$C_s=C_v/\rho$&$\kappa$&$\beta$&$V_l$&$V_t$&$\theta$&$\gamma_L$&$\gamma_T$&$\langle\gamma^2\rangle-\langle\gamma\rangle^2$&$\langle\gamma^2\rangle^{\dagger\!\dagger}$&$\tau$\\
\hline
Si\cite{duwel2011thermal}&2330&713&145\cite{grove1967physics}&2.6\cite{grove1967physics}&7470&5860&640&0.65&0.75&0.46\cite{nava1976akhiezer}&0.03\cite{nava1976akhiezer}&6.7$^\ddagger$\\
SiO$_2$&2300\cite{mitmatdatabaseSiO2}&1000\cite{mitmatdatabaseSiO2}&1.1\cite{mitmatdatabaseSiO2}&0.5\cite{madou2003fundamentals}&5748\cite{mitmatdatabaseSiO2}&3333\cite{mitmatdatabaseSiO2}&290\cite{kaviany2002principles}& $-2.1$\cite{wang2003gruneisen}&$-1.75$\cite{wang2003gruneisen}&0.88\cite{nava1976akhiezer}&0.38\cite{nava1976akhiezer}&0.13$^\ddagger$\\
SiN&2500\cite{mitmatdatabaseSiN}&170\cite{mitmatdatabaseSiN}&19\cite{madou2003fundamentals}&0.8\cite{madou2003fundamentals}&8788\cite{mitmatdatabaseSiN}&5053\cite{mitmatdatabaseSiN}&290\cite{morelli2002thermal}&0.4\cite{bruls2001temperature}$^\dagger$&0.4$^\dagger$&--&--&4.2$^\ddagger$\\	
\hline
\end{tabular}}
\caption{Material properties used for calculating Q limits due to thermoelastic and phonon-phonon dissipations at $T = 300K$. $^\dagger$ Bulk value for $\beta$-Si$_3$N$_4$ used as estimate. $^\ddagger$ Calculated using $\tau = \frac{3\kappa}{\rho C_sV_D^2}$\cite{braginsky1985systems,maris2012interaction}. $^{\dagger\!\dagger} \langle\gamma\rangle$ vanishes for transverse waves \cite{lewis1968attenuation,mason1966crystal}.}\label{tab:material}
\end{table}
\twocolumngrid

Using material property values listed in Table~\ref{tab:material}, Fig.~\ref{fig:Qlimit}(a) plots the upper bounds, due to each of TED and PPD, to the quality factors (Q) achievable in Si, SiO$_2$ and SiN, at room temperature (T = 300K).  The upper bound Q is a loss Q of a resonator implemented in the respective material, and a specific acoustic resonator design based on evanescent confinement using multiple materials will see a average of these losses with the mode field as weighting function, in addition to possible interface losses, but these values provide bounds.  Since there is very limited data in the literature on $\langle\gamma\rangle$ and $\langle\gamma^2\rangle$ for materials other than Si, a slightly different form of the $Q_\text{PPD}$ in the low frequency Akhieser regime ($\omega\tau\ll1$) for longitudinal waves was plotted in addition to Eq.~\ref{eqn:PPD}, following derivation from Duwel \emph{et al.}\cite{duwel2011thermal}
\begin{equation}
Q_\text{AKE,L} = \dfrac{\rho V_l^5\hbar^3}{k_B^4T^4}
\left[\dfrac{3\omega\tau}{2\pi^2}\left(\dfrac{V_l}{V_t}\right)^3\int_0^{\nicefrac\theta T}\dfrac{x^4e^xdx}{(e^x-1)^2}\right]^{-1}
\gamma_\text{eff}^{-2}\label{eqn:AKE}
\end{equation}
where $\theta$ is the Debye temperature. Eqn.~\ref{eqn:AKE} is evaluated in addition to Eqn.~\ref{eqn:PPD} for comparison, using effective Gr{\"u}neisen parameter $\gamma_\text{eff} = (\gamma_L+2\gamma_T)/3$, where $\gamma_L$ and $\gamma_T$ are the longitudinal and transverse Gr{\"u}neisen parameters \cite{wang2003gruneisen,anderson2000gruneisen}.  
$Q_\text{TED}$ drops in inverse proportion to frequency, as does $Q_\text{PPD}$ in the Akhieser regime. As $\omega\tau$ exceeds $1$, $Q_\text{PPD}$ turns around and improves as frequency increases. For all three materials (and as a more general rule of thumb), PPD is the more limiting loss mechanism.

A key interest in phononics is as a competing or complementary and potentially synergistic technology to photonics in signal processing, where relevant context is provided by assessing signal processing bandwidths, or potential interaction with optical modes via radiation pressure, photoelasticity or electrostriction.  Although the Q limits in Fig.~\ref{fig:Qlimit}(a) for acoustic resonators may appear low compared to optical devices ($Q \equiv \omega_o \tau_r/2$, with $\tau_r$ the lifetime), the critical metric is the resonance lifetime $\tau_r$ and associated linewidth $2/\tau_r$, and the much lower center frequency of acoustic waves makes it possible to design resonant devices with much narrower linewidths than typical optical devices on chip that are compatible with silicon processing and usually support no narrower than several GHz of bandwidth.  The dashed lines in Fig.~\ref{fig:Qlimit}(b) are calculated resonance linewidths that correspond to the material loss limited $Q$ in Fig.~\ref{fig:Qlimit}(a). Pure shear wave wavelength in bulk Si is plotted versus frequency in solid line to provide a characteristic length scale [shear wave slowness in Si cladding cuts off guided modes in the all cladded waveguide geometries, c.f. Fig.~\ref{fig:modes}(d)]. For modern fabrication processes, a relevant range of device dimensions is between $100$nm and $10\mu$m, corresponding to operating frequencies in the $1-100$ GHz range and material intrinsic linewidths from $1$\,kHz to $10$\,MHz in SiO$_2$. In terms of spatial propagation, these linewidths correspond to a loss Q of about 3,000 to 300,000, which can be converted to an equivalent waveguide propagation loss attenuation constant in dB/mm by the relation $\alpha_\mathrm{dB/mm} = \frac{0.01}{\ln(10)} \frac{\omega}{2 Q v_g}$.  The approximate attenuations corresponding to the gray region in Fig.~\ref{fig:Qlimit}(b) are 0.004 to 90 dB/mm (assuming $v_g \approx 3,000$\,m/s, i.e. SiO$_2$ core).  With compact circuits down $10-100$\,$\mu$m this is still sufficient for useful functions even at the higher end of the frequency range.  These linewidths and lengthscales suggests a potentially viable technology for discriminating signals of such or larger bandwidths and/or producing delays (and associated propagation distances) about the inverse of these bandwidths with reasonable loss.

Alternatively to Eqns.~\ref{eqn:PPD} and \ref{eqn:AKE} and similar expressions where acoustic attenuation due to PPD is evaluated using Gr{\"u}neisen parameters, the effective phonon viscosity method can be used to describe material loss from the same mechanism \cite{lamb1959absorption,helme1978phonon,wolff2014germanium,li2016attenuation}.  The phonon viscosity tensor is a representation that relates stress $\mathbf T$ to strain $\mathbf S$ as $\mathbf T = \mathbf c:\mathbf S + \boldsymbol \eta:(\partial_t \mathbf S)$. It is analogous to the imaginary part of complex refractive index (i.e. permittivity tensor) for optical waves, which describes material absorption. The effective attenuation constant $\alpha$ of a particular mode of an acoustic waveguide can be calculated by an overlap integral of the displacement field $\mathbf u$ with the material viscosity tensor $\boldsymbol \eta$, 
$\alpha = \frac{\omega^2}{P_B}\int d^2r\sum_{jkl}u_i^*\partial_j\eta_{ijkl}\partial_k u_l$, where $P_B$ is the power of the acoustic mode \cite{wolff2014germanium}. 
While an analogue to optical refractive index provides an intuitive formalism and straightforward accounting of mode field distribution in overall loss via overlap integrals, there are discrepancies between the experimental values of the viscosity tensor of Si as measured in the few available previously published experiments \cite{li2016attenuation}.  Further, the viscosity tensors of SiO$_2$ and SiN do not appear to be available in existing literature \cite{wolff2014germanium}.  For these reasons the Akhiezer model with Gr{\"u}neisen parameters is used in this paper to provide a rough estimate of limitations on the acoustic loss Q due to phonon-phonon interactions, which is sufficient for evaluating the promise of wavelength-scale phononic components and multi-element ``circuits''.  A more accurate characterization of the bulk material losses, as well as of material interface losses which we do not address here, could become important in the detailed design of devices but are beyond scope for our discussion.  

\begin{figure}[!htbp]
\includegraphics[width=.5\textwidth]{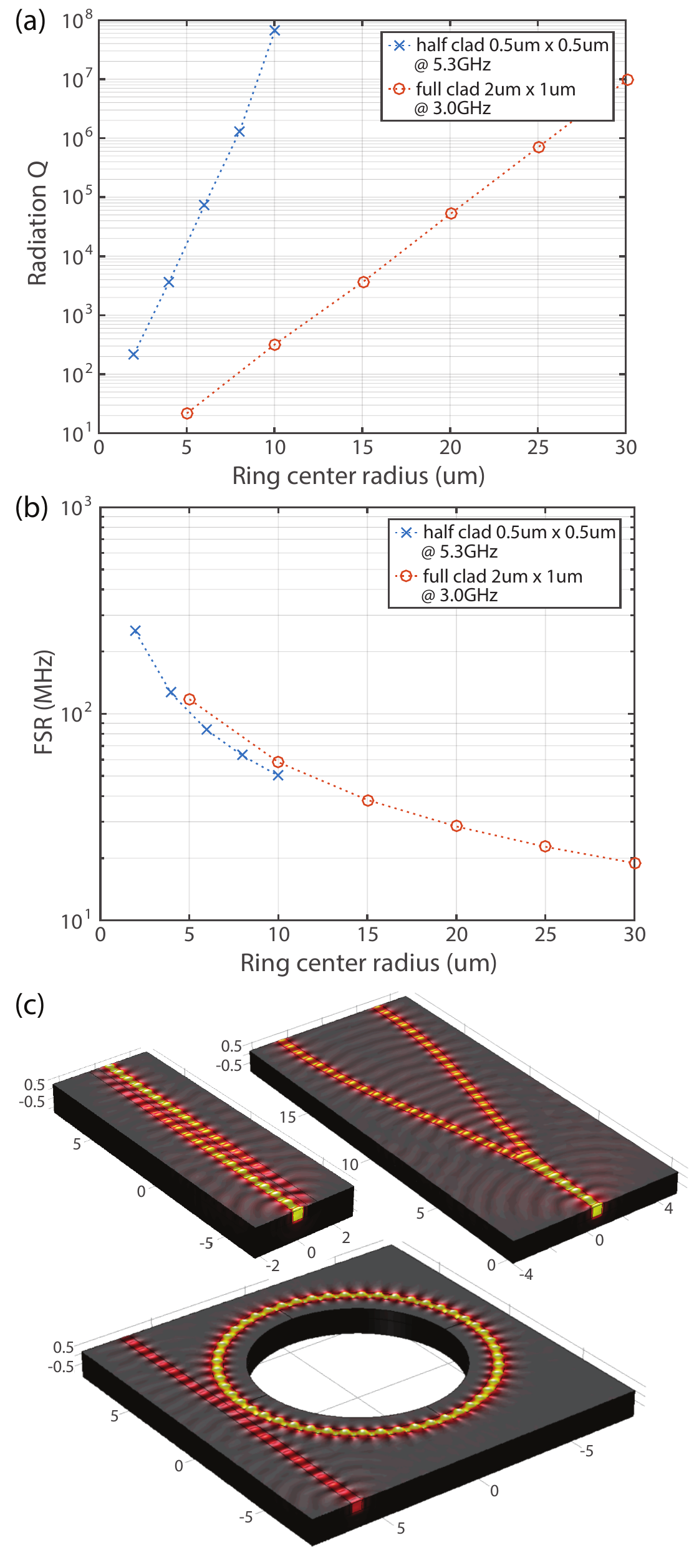}
\caption{(a) Radiation loss Q and (b) free spectral range versus ring center radius of half-cladded rings with $0.5\mu$m$\times 0.5\mu$m SiO$_2$ core and fully-cladded rings with $1\mu$m$\times 2\mu$m SiO$_2$ core from finite element simulations. (c) Simulated field profiles of acoustic directional coupler, Y-splitter and coupled ring resonator using an embedded waveguide cross-section with $0.5\mu$m$\times 0.5\mu$m SiO$_2$ core half-cladded with Si.}
\label{fig:components}
\end{figure}

\section{Radiation Loss and Design of Microphononic Circuit Components}
\label{sec:components}

In this section, we discuss the design of components.  The intrinsic material losses set the upper limits for propagation length and time delay, and complex circuits are only possible if useful functions can be accomplished in smaller length and time scales, thereby allowing several components to be traversed before the signal is lost.  We show here that high slowness contrast allows compact enough components and strong enough confinement to enable practical microphononic circuits, including coupling and routing of signals between elements.  A second key loss mechanism which is critical to determining the compactness of components (and hence viability of the circuits) is radiation loss, a generalized form of anchor loss that we address first.

The evanescent confinement between fast and slow materials enables a family of guided, traveling acoustic wave components that can be designed using techniques similar to those developed for photonic components.  Fig.~\ref{fig:components} shows simulations of example components we designed: an evanescent directional power coupler, a 1$\times$2 3\,dB wave power splitter, and an acoustic microring resonator.  They are based on a 0.5$\times$0.5\,$\mu$m silica core embedded into a silicon substrate, similar to Fig.~\ref{fig:halfclad}.  We simulated guided acoustic wave propagation through these devices, and find that compact components on the order of 10\,$\mu$m in dimension can achieve full power transfer, power splitting, or provide a functioning high-Q resonator.  We next discuss design in more detail.

Radiation loss occurs when a confined mode has an accessible radiation channel, and loses energy (referred to as a leaky mode).  Straight embedded acoustic waveguides can be designed to be fully confining, with no radiation loss.  Curved acoustic waveguides formed in a solid, however, have a fundamental radiation loss mechanism, analogous to that in optical waveguides.  Because there is an evanescent tail of the acoustic wave extending into the cladding material (orthogonal to the mode propagation direction), and the phase fronts circulate azimuthally around a circular waveguide bend, there is a radius at which the guided wave phase fronts exceed the local speed of sound.  This radius defines the acoustic radiation caustic, and results in radiation.  The radiation loss increases exponentially with smaller radius of curvature of the waveguide.  This loss mechanism limits how small a radius can be used to form a waveguide bend to route an acoustic signal between components, or an acoustic microring resonator.  Fig.~\ref{fig:components}(a) plots the resonator Q due to bending-induced radiation loss for two designs of acoustic ring resonator, as a function of the ring resonator radius.  One can also obtain from these curves the single-pass signal attenuation in dB per 90-degree turn due to bending, relevant to compact routing of signals between components, as L$_\mathrm{dB90} = \frac5{2 \ln 10} \frac{2 \pi R\omega }{v_g Q}$.  

In Fig.~\ref{fig:components}(a), one ring design has a half-cladded cross-section (see Fig.~\ref{fig:modes}(b)[iii], dimensions in caption of Fig.~\ref{fig:components}), and is designed for a resonant frequency of 5.3\,GHz.  As the radius is varied, different azimuthal mode orders (number of wavelengths around the ring) correspond to the 5.3\,GHz resonance frequency.  A very small increase in radius (to a few microns) is needed to make the radiation loss negligible, and the total Q to be limited by another (e.g. intrinsic material loss) process.  A second ring design uses a fully-cladded cross-section (Fig.~\ref{fig:modes}(b)[i], dimensions in caption of Fig.~\ref{fig:components}), and is designed for a 3.0\,GHz resonant frequency at various radii.  The first design is shown in Fig.~\ref{fig:components}(c)[iii], for a radius (to center of ring waveguide) of 6\,$\mu$m, coupled to a bus waveguide.  The bus waveguide excited the ring resonator via evanescent coupling, i.e. same power transfer mechanism seen in the directional coupler in Figure~\ref{fig:components}(c)[i], discussed next.  This provides a 2-port (notch, or all-pass) filter function.  Both ring resonator geometries allow for tight micron scale bends that are suitable for integrated systems on chip.

Because each azimuthal order yields a resonance, 
the ring resonator has a period pattern of resonances with a spacing called the free spectral range (FSR).  Smaller-radius resonators have fewer wavelengths around, so they need a larger increase in frequency to add a full extra wavelength and reach the next resonance condition -- that is, the spacing between azimuthal mode frequencies is larger.  The FSR is given by the inverse of the round trip travel time (group delay) around the ring, i.e. FSR$ = v_g/L_\mathrm{rt}$, where $L_\mathrm{rt}$ is the cavity round trip length.

Figure~\ref{fig:components}(b) shows the free spectral range (FSR) of the two specific designs showing that several MHz to several tens of MHz FSRs are supported.
The FSR can accommodate a number of frequency channels and serve as a multiplexer if the FSR is much larger than a passband of one filter.  Assuming radii are chosen large enough, bending loss Q's can exceed 10$^4$ to 10$^5$, so that intrinsic linewidths are 0.05\,MHz to 0.5\,MHz.  The other loss mechanism considered here was material intrinsic loss, and Fig.~\ref{fig:Qlimit} shows that at $3-5.3$\,GHz frequencies the intrinsic linewidth is limited to $20-70$ kHz, corresponding to Q's of 75,000 to 150,000.  Total loss Q is due to the summation of inverse Q's, $1/Q_\mathrm{total} = 1/Q_\mathrm{intrinsic} + 1/Q_\mathrm{bending}$. Thus, total loss Q's in the 10$^4$ to 10$^5$ range might be expected, giving intrinsic cavity linewidths of 50\,kHz to 500\,kHz.  A resonant filter of the kind shown in Fig.~\ref{fig:components}(c)[iii], configured with a waveguide-ring gap spacing to produce critical coupling results in a notch filter with twice the bandwidth, 100\,kHz to 1\,MHz.  Thus, with several MHz FSR, a frequency demultiplexer comprising several channels to tens of channels could be designed.  In general, since radiation loss is subject to design (e.g. choice of ring radius) while material losses are more constraining
, a designer will usually aim for a radiation Q that considerably exceeds the intrinsic Q so as to not further degrade the loss Q and linewidth.  This analysis shows that compact, few-micron-scale bends and resonators are realizable in an embedded acoustic waveguide platform with  high slowness contrast.  
Beyond ring resonators, other acoustic components can be designed as well, by analogy with their optical counterparts.

Having covered the viability of waveguides and resonators, the two basic elements of wave systems in that one carries power and the other stores energy, we next turn to the fundamental elements needed to interconnect them.  Our goal is again to evaluate the scaling of these components in the high slowness contrast regime.  Fig~\ref{fig:components}(c) shows field profiles from frequency domain simulations of an acoustic directional coupler [left] and a Y-splitter [middle], using an embedded waveguide cross-section with $0.5\mu$m$\times 0.5\mu$m SiO$_2$ core half-cladded with Si (c.f. Fig.~\ref{fig:modes}b [iii]).

The directional coupler (DC) is a fundamental component, having two input ports and two output ports, that enables a designed splitting ratio of the power in a wave at one input port into the two output ports.  DCs enable the connection of resonators to ports as in the example given above, and the construction of interferometers.  The DC shown in Fig~\ref{fig:components}(c)[left] employs evanescent coupling whereby modes of two waveguides that are in closer transverse proximity than the extent of their evanescent fields outside the core will interact and exchange significant power if they are synchronous, i.e. their propagation constants are matched, which automatically occurs with two identical guides.  A few comments can be made with respect to the scaling of directional couplers.  The basic figure of merit for a directional coupler is the length for a given fraction of power transferred -- here we choose the full-power transfer length as a baseline.  In the high slowness contrast regime, the power transfer can be rapid and full-transfer lengths very short -- order of 10\,$\mu$m in the case of Fig.~\ref{fig:components}(c)[left].  Since power transfer from one waveguide to the other after length $l$ of coupling is $|t_{21}|^2 = \sin(\kappa l)^2$, where $\kappa$ is the coupling strength (in rad/m) which falls exponentially with the gap between the waveguides due to the exponential evanescent field.  The full power transfer length, or beat length, is given by $l_\mathrm{full} = \pi/(2 \kappa)$, and $\kappa$ can be related to the propagation constants of the symmetric and antisymmetric guided supermodes of the guided pair, $\kappa = \beta_s - \beta_a$ -- the stronger the coupling, the higher the splitting of $\beta_s$ and $\beta_a$.  Since the phase velocities of the supermodes $v_{p,s}$ and $v_{p,a}$ typically fall between the core and cladding wave speeds for bulk dominated modes, the guidance condition and material slowness contrast imposes an upper limit on the coupling strength, $\kappa<\omega/v_{p,core} - \omega/v_{p,cladding}$. 
Clearly, high slowness contrast provides a larger upper bound to the coupling strength, and hence shorter coupling length.  For our case using SiO$_2$ core and Si cladding, and assuming shear wave dominance in the modes, we can approximately use 3,000 and 6,000 m/s as the respective velocities.  For frequencies in the gray region in Fig.~\ref{fig:Qlimit}(b) (about 1 to 100\,GHz), this lower bound on coupler length is 5\,$\mu$m or shorter depending on frequency.  Note that all of these estimates are independent of particular geometry.  In reality confinement and design for radiation loss will produce designs with longer lengths -- consistent with our 10\,$\mu$m length example in Fig~\ref{fig:components}(c)[left].  Directional couplers can be straight, and do not incur limitations of bending losses, although bringing isolated waveguides to them does, so the total length of a coupler including connections might be 2-3 times the size given our estimates of bend radii.

Another fundamental component of wave circuits (such as integrated optical circuits, or microwave circuits) are Y-branch 3\,dB splitters.  Directional couplers provide arbitrary splitting ratios, but broadband designs are difficult to achieve even in integrated optics, where 1\% bandwidth is considered broadband (i.e. 2\,THz of a 200\,THz carrier).  In acoustic circuits, one may desire 10\% or higher bandwidths, requiring wideband design more akin to microwave engineering than integrated photonics. DCs are furthermore sensitive to fabrication variations.  Hence, the Y-branch splitter is a device that guarantees 3\,dB (50:50) splitting, an important ratio for interferometers and power splitter trees, by symmetry.  The Y-branch splitter involves a splitting region and branch arms to separate the ports.  The splitting region can be very compact -- a few microns in both in-plane dimensions following similar slowness contrast arguments to the DC length.  Based on bend radii of order 10\,$\mu$m, Fig~\ref{fig:components}(c)[middle] shows a 3\,dB power splitter that using S-bends that is about 20\,$\mu$m long and 10\,$\mu$m wide (including port separation) for 3.35\,GHz operating frequency.  Smaller structures are possible using multi-mode interference or more advanced taper concepts and more aggressive bend design.

\section{Conclusion}
Microphononic circuits can realize both coupled-element circuits via evanescent coupling based on embedded waveguides (analogously to dielectric integrated photonic circuits), or via physically interconnected suspended structures (closer to microwave circuits based on metal-walled microwave cavities, where the free boundary in acoustics corresponds to the perfect electric conductor wall of a microwave cavity).  In this paper, we investigated the former and found that the confinement and losses are consistent with enabling micron-scale devices and circuits that operate on 1-100\,GHz bandwidth signals, where the low end of frequencies is likely to be limited by device size and the high end by losses and lithographic resolution.

High-slowness-contrast embedded acoustic waveguides allow complete, radiation-free guided wave confinement in 1\,$\mu$m scale cross-sections.  Intrinsic material losses permit operation in the 1-100\,GHz frequency range with Q's in the thousands and linewidths that permit efficient signal processing -- and tens of microns to millimeters or centimeters of low loss propagation depending on frequency.

Basic building blocks based on evanescent confinement and coupling of embedded structures include waveguide bends, ring resonators, directional couplers and Y splitters that all benefit from the high slowness contrast to provide 10\,$\mu$m scale structures at few-GHz frequencies, thus enabling complex multi-element circuits.  We believe this kind of microphononic circuit platform, which does not require suspended components, warrants further investigation, and could provide valuable components integrable directly with microelectronics and microphotonics in CMOS as well as specialized custom chip technology.

Evanescently confined microphononic circuits could find applications in sensing, communication and RF signal processing, interfaced either to electronics or photonics.  With additional attention paid to simultaneous confinement of optical and acoustic waves, applications in optomechanics, signal domain transduction, and microwave photonic signal processing could benefit.  The presented study of guidance and loss supports the feasibility of such structures in dimension and frequency ranges of interest, and suggests the next steps of experimentally demonstrating these geometries.  The possibility of microphononic circuits being incorporated within planar CMOS technology, could enable complex systems-on-chip that can benefit from the narrowband signal processing capabilities and long time delays enabled by confined sound waves.  Investigation of some of these possibilities is a worthy subject for future work.

\begin{acknowledgments}
This work was supported by a 2012 Packard Fellowship for Science and Engineering (Grant $\#$2012-38222).
\end{acknowledgments}
\end{document}